# Photoconductance of a submicron oxidized line in surface conductive single crystalline diamond

M. Stallhofer, M. Seifert, M. Hauf, G. Abstreiter, M. Stutzmann, J. Garrido, and A.W. Holleitner[a)]

Walter Schottky Institut and Physik Department, Technische Universität München, Am Coulombwall 3, D-85748 Garching, Germany.

We report on sub-bandgap optoelectronic phenomena of hydrogen-terminated diamond patterned with a submicron oxidized line. The line acts as an energy barrier for the two-dimensional hole gas located below the hydrogenated diamond surface. A photoconductive gain of the hole conductivity across the barrier is measured for sub-bandgap illumination. The findings are consistent with photogenerated electrons being trapped in defect levels within the barrier. We discuss the spatial and energetic characteristics of the optoelectronic phenomena, as well as possible photocurrent effects.

a) corresponding author: holleitner@wsi.tum.de

Diamond spurs scientific and technological interest because of its high thermal conductivity, a large dielectric breakdown, and a large bandgap of ~5.5 eV.[1] Further functionality can be achieved by utilizing undoped single crystalline diamond with a hydrogen termination, which exhibits a surface conductivity in air.[2] The surface conductivity originates from a two-dimensional hole gas that is formed due to an upward bending of the valence band at the hydrogenated surface.[3,4,5,6] The extent of the hole gas is in the range of a few nm, depending on the quality of the hydrogen termination.[4] Recently, the surface conductive layer has been exploited to realize in-plane gated field effect transistors.[7,8,9,10] In addition, hydrogenated diamond surfaces exhibit a negative electron affinity.[11,12] The negative electron affinity becomes positive when the surface is oxidized. The change in electron affinity leads to a depletion of the hole gas. Patterning a thin oxidized line into hydrogenated areas of a diamond surface results in a lateral energy barrier for the two-dimensional hole gas.[8,9] The corresponding band alignment has been recently verified by Kelvin force microscopy, and it is sketched in Fig. 1(a).[13,14,15,16]

Here, we investigate the sub-bandgap optoelectronic phenomena induced by such a barrier in hydrogenated diamond at room temperature. The submicron lines are defined in single-crystalline diamond by electron beam lithography in combination with an oxygen plasma treatment. We find a photoconductive gain (photo-transistor) effect,[17,18,19,20] which is explained by the influence of photogenerated electrons trapped in defect states located in the oxidized lines [filled circle in Fig. 1(a)]. The energy of the defect levels, which cause the photoconductive gain, is found to be at (2.6 ± 0.2) eV above the valence band. We observe a typical response time of the presented diamond based circuits in the order of a few hundreds of milliseconds. At large laser power, a space charge limited current across the barrier dominates the photoconductance.[21] Our findings demonstrate that surface conductive diamond circuits can be tailored by submicron oxidized lines in order to build photodetectors in the ultra-violet range at room temperature.

The experiments are performed using an electronic grade, chemical vapor deposition (CVD)-grown single-crystalline diamond with [100] orientation.[22] The surface conductivity is created by exposing the surface to a hydrogen plasma.[23] The samples typically show a hole density of $10^{13}$ $cm^{-2}$ and a carrier mobility of 50 to 100 $cm^2/Vs$. Ohmic contacts to the conducting layer are achieved by a lateral overlap of gold pads with the conductive channels.[24] The submicron lines are defined by e-beam lithography,[25] and the oxidation of the lines is done in an oxygen plasma at 200 W for 180 s. We present results of devices with lines exhibiting a lithographic width of 70 nm and 1 μm. All measurements are carried out at room temperature in a vacuum of $\sim 10^{-5}$ mbar. Without optical excitation, the resistance between two Ohmic contacts with an oxidized line in between exceeds ~200 GΩ. We explain this large dark resistance by the potential difference of ~2.7 eV between the p-type hydrogenated surface (with a Fermi-energy $E_{Fermi}$ of 0.7 eV below the valence band) and the oxidized line (with the valence band ~2 eV below $E_{Fermi}$) [Fig. 1(a)].[7,9,15]

Optical excitation occurs by focusing the light of a mode-locked titanium:sapphire laser with a repetition rate of 76 MHz through the objective of a microscope onto the diamond circuits. In combination with a beta-$BaB_2O_4$ (BBO) crystal and a photonic glass fiber, the laser excitation can be continuously tuned in the energy range of 1.24 eV < $E_{photon}$ < 3.35 eV. With a spot diameter of 2.9 μm, the light intensity $I_{laser}$ is in the order of $10^2$ W/cm$^2$ for all $E_{photon}$. We chop the laser at a frequency $f_{chop}$ which is varied between 1 Hz and 1 kHz. The resulting current

$$I_{photo} = I_{photo}^{ON}(E_{photon}, f_{chop}) - I_{photo}^{OFF}(E_{photon}, f_{chop}) \quad (1)$$

across the sample with the laser being "on" or "off," respectively, is amplified by a current-voltage converter and detected with a lock-in amplifier utilizing the reference signal provided by the chopper.

A typical spatial dependence of $I_{photo}$ is depicted in Fig. 1(b), when the laser is scanned across an oxidized line with a lithographic width of 1 μm. We find a maximum of the optoelectronic response at the position of the oxidized line [triangle in Fig. 1(b)]. Fig. 2(a) depicts a full map of $I_{photo}$ as a function of the coordinates $x$ and $y$ for the same line as in Fig. 1(b). Bias dependent measurements demonstrate that $I_{photo}$ is symmetric with respect to zero $V_{sd}$ [triangles in Fig. 2(b)].[26] The bias-symmetry at any coordinate [data not shown] suggests that the optoelectronic signal is dominated by a photoconductance effect and not by a photovoltaic effect.[17,18,19,27]

Oxidized diamond is known to exhibit surface defect states.[7] Hereby, the optoelectronic observations can be interpreted as follows. After a sub-bandgap illumination, electrons get excited into traps within the bandgap [vertical arrow in Fig. 1(a)]. The trapped electrons affect the electrostatic potential landscape of the circuit such that they lower the barrier for holes centered at the position of the oxidized line. In turn, an increased optically induced conductance is measured. Most importantly, such a photoconductance is determined by the occupation of the defect states.

This interpretation is supported by a quadratic dependence of $I_{photo}$ on $V_{sd}$ at a large laser power $P_{laser}$ [circles and line in Fig. 2(b)], for the laser being focused onto the center of an oxidized line [triangle in Fig. 1(b)]. The quadratic behavior is characteristic of a space charge limited current.[21] In this scenario, photogenerated charge carriers accumulate around the point of creation. The Coulomb repulsion or a saturation of the absorption hinders further electrons to be excited into the defect states in the oxidized lines. At small $P_{laser}$, however, the reduced optical generation rate leads to a linear response at small $V_{SD}$ [triangles in Fig. 2(b)]. When the laser is not focused onto the oxidized line [as exemplarily marked by the diamond symbol in Fig. 1(b)], we observe a linear $I_{photo}$ - $V_{sd}$ characteristic also at a large $P_{laser}$ [diamond symbols in Fig. 2(c)], which is tentatively attributed to the photoconductance of hydrogenated diamond.[28] This is verified by plotting the spatial dependence of the phase signal $\phi$ of the lock-in measurement at zero $V_{sd}$.

Fig. 2(d) shows such data for the laser being scanned across a 70 nm line. As can be seen, the phase in the areas on the left and right side of the oxidized barrier differs by 180° with a transition at the position of the barrier. Generally, the source- and drain-reservoirs are coupled capacitively across the oxidized line. In turn, we can detect an optically induced displacement current $I_{photo}$ by the lock-in measurements far away from the line, although a non-illuminated line blocks the dc-current from source to drain. Depending on the position of the laser relative to the oxidized line, the photogenerated hole density increases the chemical potential of either the source- or the drain-reservoir.[17] This explains the phase-jump of 180° in Fig. 2(d). An alternative interpretation considers stray light that induces the photo-transistor effect at the oxidized line, although the laser spot is located far away from the line. However, when measuring $I_{photo}^{ON}$ without a lock-in, we cannot resolve a signal above the noise level (data not shown). Therefore, the alternative explanation seems less likely.

Generally, there also should be a photocurrent due to holes photogenerated in the vicinity of the oxidized line.[20] These holes propagate either to the source or the drain reservoir [horizontal arrows in Fig. 1(a)]. At present, we cannot exclude that the phase-change in the direct vicinity of an oxidized line, as in Fig. 2(d), also comprises the contribution of such a p-type photocurrent. The spatial FWHM of the optoelectronic response of an oxidized line is several micrometers, and it is much larger than the lithographically defined widths. We interpret the FWHM to originate from a space charge distribution with micrometer extension, which is consistent with recent results from Kelvin force measurements.[13,14,15,16]

In the following, we address the energetic position of the defect states within the oxidized barrier. Fig. 3(a) shows $I_{photo}$ as a function of $E_{photon}$, when the center of an oxidized line is excited. We observe an onset of the photoresponse at $E_{photon} = 2.6$ eV ± 0.2 eV. The value compares reasonably well with the ones reported in literature for defect states in oxidized diamond surfaces.[3,8,28] In this picture, the relatively broad width of the transition of ±0.2 eV reflects the spectral width across the spatial distribution of these levels at the oxidized line. This observation is consistent with the occurrence of a space charge limited current at a large $P_{laser}$ as discussed above. We further interpret the saturation of the curve in Fig. 3(a) for $E_{Photon} > 2.8$ eV such that there are no further resonant energy levels above this energy. The defect states are different to the ones induced by nitrogen impurities that lie 1.8 to 2.2 eV below the conduction band, and which have been recently exploited for optically induced transport measurements.[29].

We point out that we observe an optoelectronic response time of around hundreds of milliseconds in the described diamond based circuits. Fig. 3(b) exemplarily depicts the decay time of the optically induced current $I_{photo}^{ON}$ at the center of an oxidized line, after blocking the laser. Here, $I_{photo}^{ON}$ is measured in a time-integrated way without using a lock-in. Such a measurement yields the large signal response time of the diamond based circuit. Fitting the data to a mono-exponential decay function gives an optoelectronic response time of (590 ± 20) ms [line in Fig. 3(b)]. This is distinctively shorter than the characteristic decay time of ~3 hours found for nitrogen defect states in diamond.[29] In our case, the time-scale is probably due to hopping of electrons from higher defect levels located in the transition region between the oxidized and the hydrogenated surface, to the energetically favorable states within the oxidized barrier. These electrons contribute then to the reduction of the potential barrier, before they recombine with holes. We note that the so-called low level signal $I_{photo}$, as defined in Eq. (1), is independent of $f_{chop}$ for 1 Hz $< f_{chop} <$ 1 kHz within the experimental error.

We further find that $I_{photo}$ depends linearly on $P_{laser}$ before the space charge limited current sets in at a large $P_{laser}$. Thus, we can exclude two-photon processes to explain the reported optoelectronic characteristics of the devices. Instead, the linear dependence is consistent with a dominating photoconductance effect as discussed above. In addition, the chosen lithographic width of the oxidized lines of up to 1 μm makes internal photo-emission processes very unlikely to contribute to $I_{photo}$.

In summary, we present optoelectronic phenomena in hydrogenated diamond caused by a submicron oxidized line. We interpret our findings by a photo-transistor effect which is induced by electrons optically excited into defect levels inside the oxidized region. The presented results may prove useful to design spatially resolved photodetectors in the ultra-violet range with submicron scale resolution that operate at room temperature.

We gratefully acknowledge financial support by the DFG via Ho 3324/2 and the German excellence initiative via the "Nanosystems Initiative Munich (NIM)".

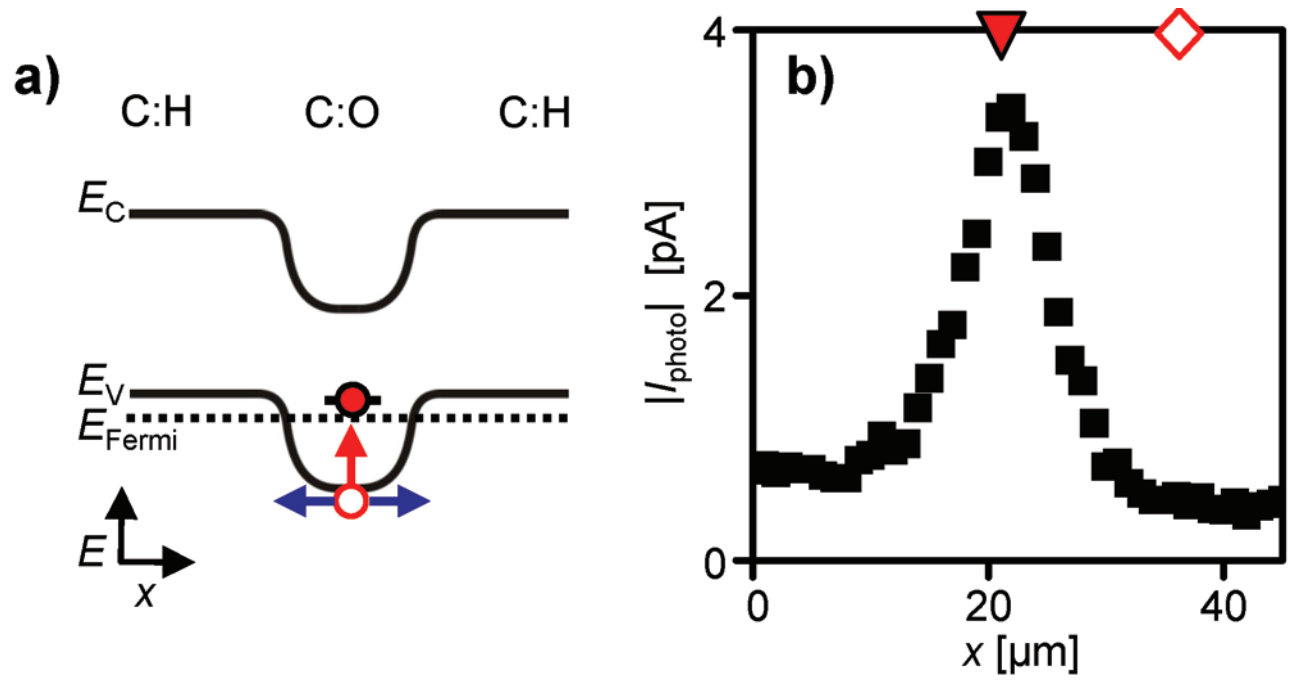

FIG. 1. (Color online) (a) Schematic view of the lateral band alignment of an oxidized line in between two hydrogenated diamond surface areas. The dotted line depicts the Fermi-energy of the two-dimensional hole gas. Sub-bandgap optical excitation (vertical arrow) results in electrons being trapped in defect states (filled circle). (b) Conductive photoresponse of the hole gas along the x-direction across such a barrier at room temperature. Triangle (diamond symbol) depicts the position of an oxidized line (a position next to it) with a lithographic width of 1 µm ($V_{sd}$ = 200 mV, $P_{laser}$ = 10 µW, $E_{photon}$ = 3.1 eV, $f_{chop}$ = 617 Hz).

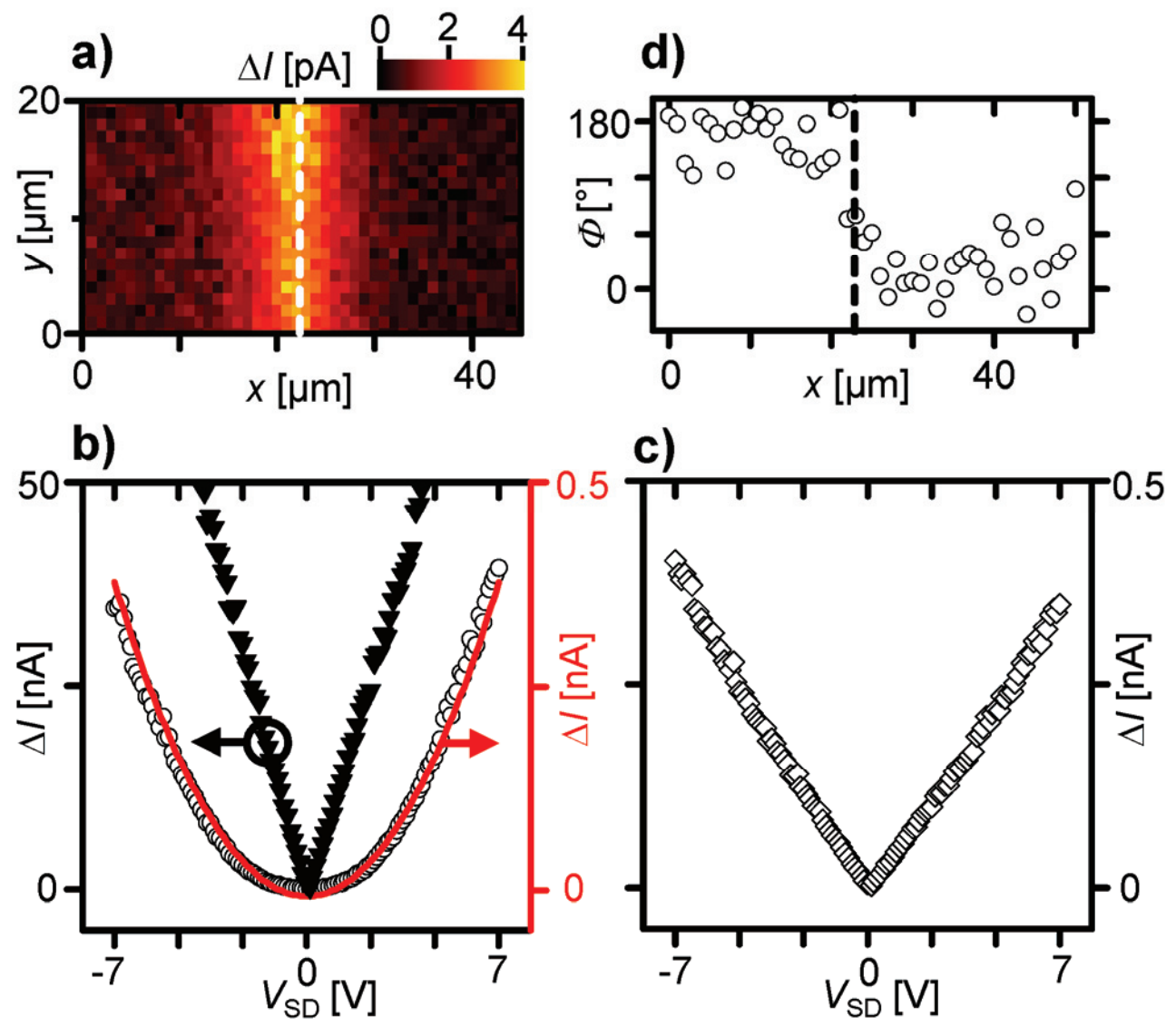

FIG. 2. (Color online) a) Optoelectronic response-map of an oxidized 1 μm wide line (marked as dashed line) between two hydrogenated diamond surfaces ($E_{photon}$ = 3.1 eV, $V_{sd}$ = 0.24 V, $P_{laser}$ = 10 μW, $f_{chop}$ = 917 Hz). b) $I_{photo}$ as a function of $V_{sd}$ with laser-spot as marked by triangle in Fig. 1(b) at $P_{laser}$ = 20 μW (triangles) and 200 μW (circles). Line is a quadratic fit ($f_{chop}$ = 77 Hz, $E_{photon}$ = 3.1 eV). c) $I_{photo}$ as a function of $V_{sd}$ with laser-spot as marked by diamond symbol in Fig. 1(b) ($f_{chop}$ = 77 Hz, $E_{photon}$ = 3.1 eV, $P_{laser}$ = 200 μW). d) Phase $\Phi$ of the lock-in signal as a function of the *x*-direction across a 70 nm wide line (position indicated by dashed line) ($E_{photon}$ = 3.1 eV, $V_{sd}$ = 0V, $P_{laser}$ = 245 μW, $f_{chop}$ = 77 Hz).

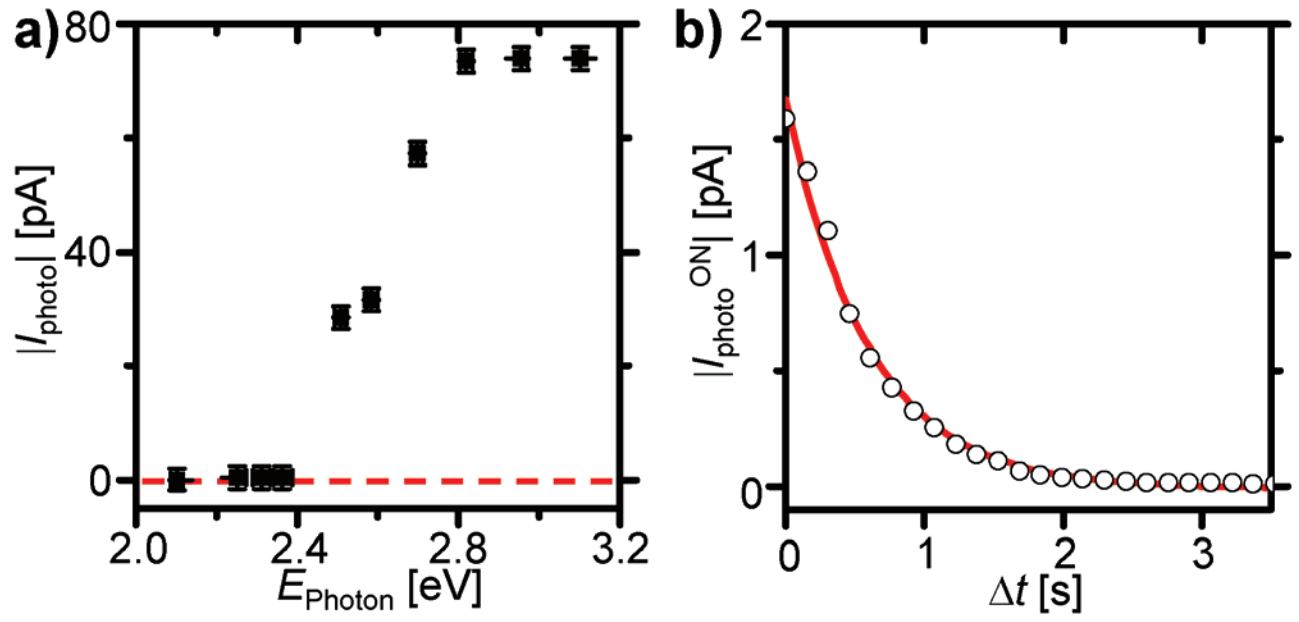

FIG. 3. (Color online) (a) Optoelectronic response across a 1 μm wide line as a function of $E_{photon}$ ($V_{sd}$ = 4 V, $P_{laser}$ = 20 μW, $f_{chop}$ = 77 Hz). b) Switch-off behavior of $I_{photo}^{ON}$ ($V_{sd}$ = 0.24 mV, $P_{laser}$ = 35 μW, $E_{photon}$ = 3.35 eV, $f_{chop}$ = 917 Hz). The line is a mono-exponential fit to the data.